\documentclass[final,5p,times,twocolumn,authoryear]{elsarticle}
\usepackage[T1]{fontenc}
\usepackage[utf8]{inputenc}

\usepackage{amssymb}
\usepackage{hyperref}
\hypersetup{
    colorlinks=true,
    linkcolor=blue,
    citecolor=blue,  
    urlcolor=blue,
    }

\usepackage{color}

\journal{Journal of High Energy Astrophysics}

\begin{document}

\begin{frontmatter}



\title{Estimation of the atmospheric absorption profile with isotropic background events observed by Imaging Atmospheric Cherenkov Telescopes}


\author[inst1]{Julian Sitarek}

\affiliation[inst1]{organization={Department of Astrophysics, University of Lodz},
            addressline={Pomorska 149}, 
            city={Lodz},
            postcode={90-236}, 
            country={Poland}}

\author[inst2]{Mario Pecimotika}
\author[inst1]{Natalia Żywucka} 
\author[inst1]{Dorota Sobczyńska}

\affiliation[inst2]{organization={Ru\dj er Bošković Institute},
            addressline={Bijenička cesta 54}, 
            city={Zagreb},
            postcode={10000}, 
            country={Croatia}}
\author[inst3]{Abelardo Moralejo}
\affiliation[inst3]{organization={Institut de Fisica d’Altes Energies },
            addressline={Campus UAB, Bellaterra}, 
            city={Barcelona},
            postcode={08193}, 
            country={Spain}}

\author[inst4]{Dario Hrupec}
\affiliation[inst4]{organization={J. J. Strossmayer University of Osijek, Department of Physics},
            addressline={Gajev trg 6}, 
            city={Osijek},
            postcode={31000}, 
            country={Croatia}}

\begin{abstract}
Atmospheric Cherenkov telescopes rely on the Earth's atmosphere as part of the detector. 
The presence of clouds affects observations and can introduce biases if not corrected for. 
Correction methods typically require an atmospheric profile, that can be measured with external atmospheric monitoring devices. 
We present a novel method for measuring the atmospheric profile using the data from Imaging Atmospheric Cherenkov telescopes directly. 
The method exploits the comparison of average longitudinal distributions of the registered Cherenkov light between clear atmosphere and cloud presence cases. 
Using Monte Carlo simulations of a subarray of four Large-Sized Telescopes of the upcoming Cherenkov Telescope Array Observatory and a simple cloud model we evaluate the accuracy of the method in determining the basic cloud parameters. 
We find that the method can reconstruct the transmission of typical clouds with an absolute accuracy of a few per cent.
For low-zenith observations, the height of the cloud centre can be reconstructed with a typical accuracy of a few hundred metres, while the geometrical thickness can be accurately reconstructed only if it is $\gtrsim3$~km.
We also evaluate the robustness of the method against the typical systematic uncertainties affecting atmospheric Cherenkov telescopes. 
\end{abstract}


\begin{highlights}
\item A novel method is proposed for the evaluation of cloud transmission profile using Cherenkov telescope data. 
\item The method provides total transmission measurement with absolute accuracy of a few per cent.
\item The cloud height is reconstructed with a small underestimation bias, while the cloud thickness can be reconstructed well only if $\gtrsim3$~km.
\item The method can serve as an auxiliary cloud parameters measurement without the need for any additional hardware or loss of observation time.  
\end{highlights}


\end{frontmatter}


\section{Introduction}
\label{sec:intro}

Imaging Atmospheric Cherenkov Telescopes (IACTs) are the main type of instrumentation used for studies of very-high-energy ($\gtrsim100$~GeV) gamma rays from cosmic sources (see e.g. \citealp{2022Galax..10...21S} for a recent review). 
A gamma ray entering the atmosphere initiates a cascade of energetic secondary particles, the so-called Extensive Air Shower (EAS).
Charged particles in an EAS, that propagate faster than the speed of light in the medium induce the emission of Cherenkov photons. 
The part of the Cherenkov radiation that is not absorbed in the atmosphere reaches the ground, where it can be registered by large telescopes (with a mirror diameter of the order of 10~m). 

The presence of clouds during observations with an IACT causes additional absorption.
Since only a part of the light is absorbed, clouds usually do not prevent the observations, however, if not taken into account, they may introduce bias in the results of the data analysis. 
Several methods have been derived to correct for the effect of clouds \citep{Nolan10f1, Hahn14f2, Devin19c1,Dorner09g1,Fruck14d2,Fruck15g2, 2020APh...12002450S,2022JPhCS2398a2011S, Schmuckermaier23e,ref:clouds}. 
Since the Cherenkov light is emitted over a range of heights above the telescope, most of the correction methods use information about the absorption profiles of the atmosphere.
Such information is typically obtained using auxiliary atmospheric monitoring devices \citep{2022JPhCS2398a2011S,Devin19c1,Fruck:2022,Schmuckermaier23e}, in particular LIDAR (LIght Detection And Ranging). 
LIDAR devices use the information from the reflected light to efficiently reconstruct the atmospheric transmission profile. 
However, the analysis method is not fully straightforward and requires additional calibrations and assumptions, especially in the case of a LIDAR that operates at a single wavelength (i.e. elastic LIDAR). 
Moreover, LIDAR devices (especially those of the Raman type) exploit powerful laser pulses, which could affect the observations, limiting their use. 

It has also been proposed to use the Cherenkov data to evaluate the transmission of the atmosphere.
For example, the comparison of the rate of events with different intensities (total registered charge) was proposed to evaluate the absorption in the Saharian Air Layer \citep{Dorner09g1}, and a combination of stereo rates and muon-derived efficiency was used to calculate the transmission coefficient of the atmosphere  \citep{Hahn14f2}.
Nevertheless, all of these methods can evaluate only the total transmission, rather than its height dependence, which is crucial due to the geometric size of the EAS in the atmosphere. 

Recently, we proposed a correction method that can use a simple geometric model to efficiently correct the IACT data already at the image level \citep{ref:clouds}. 
Similarly to methods operating at the higher level (correcting estimated energy of events and corresponding collection area), it also exploits the knowledge of the atmospheric transmission profile to derive correction factors, but at the level of individual pixels. 
In this work instead, we investigate an opposite question: can the light atmospheric profile be determined using the IACT data itself, exploiting a simple geometrical shower model? 
As an example, we consider Monte Carlo (MC) simulations of a sub-array composed of 4 Large-Sized Telescopes (LSTs, \citealp{Abe2023PerformanceData}), the largest type of IACTs forming the future Cherenkov Telescope Array Observatory (CTAO, \citealp{2013APh....43....3A}).

In Section~\ref{sec:atmprof} we explain the basic principle of the proposed method.
In Section~\ref{sec:mcs} we list the MC simulations performed.
The details of the analysis are presented in Section~\ref{sec:analysis}.
In Section~\ref{sec:results}, we evaluate the performance of the method and discuss possible systematic errors relevant to its application to real data. 
We conclude the paper with a summary and discussion in Section~\ref{sec:summary}.

\section{Derivation of the atmospheric transmission profile} 
\label{sec:atmprof}

To relate the pixels of the camera to a particular emission height we use a method very similar to \cite{ref:clouds}. 
Namely, for each event, we perform tentative stereoscopic reconstruction using the crossing point of the main axes of the ellipses. 
Then, the distance of each pixel to this point, projected at the line joining the reconstructed source position and the centre of gravity (COG) of the image is computed and, using a geometrical formula, converted into the emission height (c.f. Fig.~2 in \citet{ref:clouds}:
\begin{equation}
\xi = \arctan\left(\frac{I \cos\theta}{H}\right),
\end{equation}
where $\xi$ is the offset angle from the primary proton direction corresponding to the emission height $H$ (measured above the ground), $I$ is the (preliminary reconstructed) impact parameter, and $\theta$ represents the zenith angle of the observations (i.e. the zenith angle of the shower).

\begin{figure*}[bt!]
    \centering
    \includegraphics[width=0.9\textwidth]{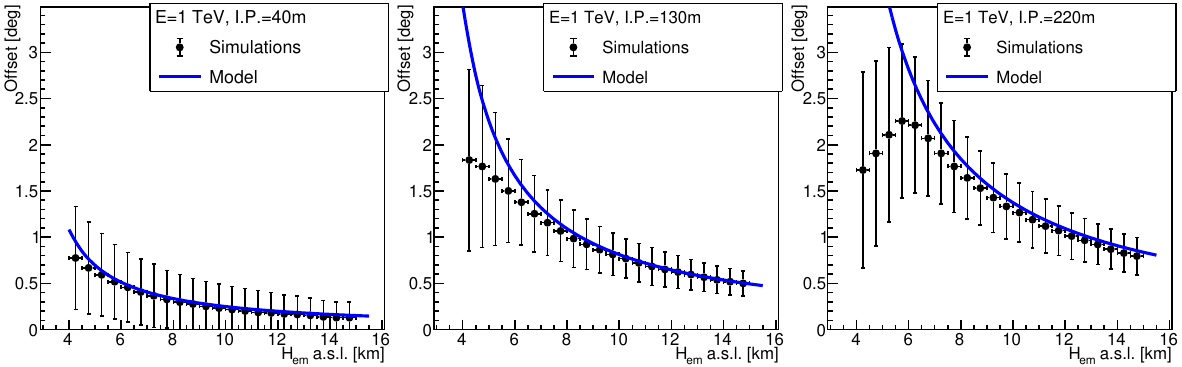}
    \includegraphics[width=0.9\textwidth]{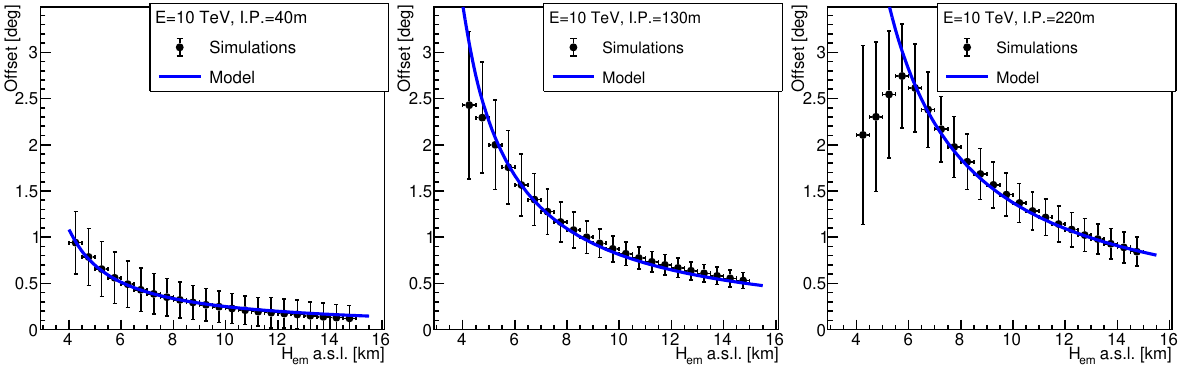}
    \caption{Offset angles of the emitted Cherenkov photons by protons with energy 1~TeV (top panels) and 10~TeV (bottom panels) observed at a zenith distance angle of $20^\circ$ as a function of the emission height for the impact parameter (I.P.) of 40~m (left), 130~m (centre), and 220~m (right).
    Points show the average offset from the shower simulations with the vertical error bars showing the standard deviation of the spread of offsets within one emission height bin. 
    Only Cherenkov photons within $\pm0.2^\circ$ from the main shower axis are considered.
    The blue solid line shows the bias-corrected geometrical model prediction according to Eq.~\ref{eq1}.
    }
    \label{fig:pmodel}
\end{figure*}

Most of the events observed by IACTs are proton-induced isotropic background.
We concentrate on the TeV range of proton energies, as we want to apply the method to large, well-reconstructed events with their emission observable from a range of altitudes. 
Similarly to \citet{ref:clouds}, we introduce a minor phenomenological correction to the geometrical model using special CORSIKA simulations of vertical proton-induced air showers observed at different energies and impact parameters, the dominant primary particle event type in the raw IACT data. The correction factor aims to improve the agreement between the model and the data since the model itself does not take into account the lateral and angular distribution of the charged particles in the air shower emitting Cherenkov photons.
Furthermore, to limit the influence of the larger perpendicular momenta of the particles in the hadronic showers we consider only the light emitted within $\pm 0.2^\circ$ from the main axis of the image.  
The corrected formula for the angular offset, $\xi'$, of Cherenkov light emitted at height $H$ (measured above the ground level) from a proton-induced shower then becomes:
\begin{equation}
    \xi' = \frac{0.85}{\cos\theta}\cdot\xi,\label{eq1}
\end{equation}
where $0.85 / \cos\theta$ is a phenomenological bias correction of the geometrical model dependent on the zenith angle $\theta$ of the observations. 
The bias-corrected geometrical model is presented in Fig.~\ref{fig:pmodel}.
The derived formula describes relatively well the full shower simulations for the studied energies of 1 and 10~TeV. 
Deviations from the model occur at high impact parameters, where the mean offset is $\gtrsim2.5^\circ$. Such events are typically clipped by the camera edge.
Compared to the case of gamma-ray-initiated showers, discussed in \citet{ref:clouds}, the spread of the offset angles of the light emitted from a given height is larger. 
Nevertheless, for large impact events and an emission height $\sim$ 10~km above ground level (a.g.l.), the spread of offsets corresponds to the spread of heights of $\sim1 - 2$~km. 
We interpret this as a natural limit to the accuracy of the method for detecting structures of geometrically thin clouds.

The proposed method for evaluating the atmospheric absorption profile of the clouds from the observations of background events is based on constructing a sum of observed longitudinal distributions of observed Cherenkov light.
Such reconstructed aggregated distributions can be then compared between clear atmosphere and cloud cases, and the ratio can be directly interpreted as the measure of the transmission of the cloud. 
In the presence of clouds, the collection area drops, and the events are in general worse reconstructed. Therefore, to compensate for the event selection bias, we renormalise the transmission profile in the 3 -- 4~km a.g.l.

\section{Monte Carlo simulations}
\label{sec:mcs}

To evaluate the performance of the proposed method, we performed Monte Carlo simulations. We used CORSIKA 7.7410 \citep{Heck:1998, Bernlohr:2008} to model air showers induced by protons arriving at a zenith angle of $20^\circ$ and an azimuth angle of $180^\circ$ (measured clockwise from the geographic North; South pointing), within a cone with a $10^\circ$ half-opening angle centred around the above zenith and azimuth angles. The UrQMD model \citep{Bass:1998ca} was used for low-energy interactions and the QGSJET-II-04 model \citep{Ostapchenko:2004qz} for high-energy interactions. The energies of the simulated events follow a power-law distribution with a spectral index of $-2$ in the energy range from 20 GeV to 300 TeV. The impact parameter was chosen such that the events were uniformly distributed within a circle with a maximum radius of $r_\mathrm{max}=1400$ m. Each shower was reused 20 times by randomly changing the impact point with respect to the nominal shower core. For each set we have simulated a total of $10^9$ air showers. 

The \texttt{sim\_telarray} \citep{Bernlohr:2008} code was used to simulate the effects of the atmosphere on the propagation of Cherenkov light (emitted in the wavelength range from 270 nm to 750 nm) and the detector response. The simulated telescope layout consists of four LSTs (see Fig. 1 in \citealp{ref:clouds}). We mainly simulate proton-induced showers. Since we also want to evaluate the robustness of the method to the typical systematic uncertainties affecting IACTs, we have produced other simulation sets as well (e.g., different zenith angles or primary particles, described in detail in Section~\ref{sec:syst}). The differences between these sets and the nominal simulations, if any, are indicated in the text.

Atmospheric transmission profiles, including clouds, were simulated using the MODerate resolution atmospheric TRANsmission (MODTRAN) code version 5.2.2 \citep{berk1987modtran,Berk2005MODTRAN5A}. MODTRAN is a radiative transfer algorithm that calculates spectral absorption, transmission, emission, and scattering in the atmosphere at moderate spectral resolution (from infrared to ultraviolet).
Various types of clouds occur at the site of the future CTA-North Observatory. \citet{Fruck:2022} has observed, using LIDAR measurements, that the majority of the clouds seen during MAGIC observations have total transmission ($T$) ranging from $\sim0.4$ to $\sim0.95$.
The base of the cloud starts at 5 km a.g.l. up to 9 km a.g.l. and the top in the range of 8 -- 11 km a.g.l.

Therefore, we consider as our ``baseline cloud'' a $T=0.587$ transmission\footnote{The value was selected to be $\approx0.6$, however, due to the non-direct way of selecting transmission in MODTRAN, it is not exactly equal to this value. Also, the effective transmission of the cloud will be decreased for non-vertical observations.} cloud spreading at a height of $H_c=6.5$ -- 8.5~km a.g.l. (with a thickness of 2~km).
The clouds are considered to be homogeneous and quasi-grey, i.e., their transmission is nearly independent of wavelength (see the discussion in \citealp{ref:clouds}).
To evaluate the performance of the proposed method, we vary the basic cloud parameters one by one.
Namely, we test clouds within a $\pm1$~km range from the base height of the baseline cloud, with the thickness of ${}_{-1}^{+2}$~km of the baseline cloud thickness, and with transmission ranging from 0.388 to 0.800.
Simulation parameters for all clouds are summarised in Table~\ref{tab:mcs}.
\begin{table}[t!]
    \centering
    \begin{tabular}{c|c|c}
        Transmission & Base height [km] & Thickness [km] \\\hline
        0.388 & 6.5 & 2 \\
        0.587 & 5.5 & 2 \\
        \textbf{0.587} & \textbf{6.5} & \textbf{2} \\
        0.587 & 6.0 & 3 \\
        0.587 & 5.5 & 4 \\
        0.587 & 7.0 & 1 \\
        0.587 & 7.5 & 2 \\
        0.800 & 6.5 & 2 \\ \hline
        1 & -- & --  
    \end{tabular}
    \caption{Parameters of simulated clouds: transmission, height of the base of the cloud above telescope level, total geometrical thickness (height of the top of the cloud minus height of the base).
    The ``baseline'' cloud is marked in bold. 
    The last row corresponds to cloudless sky. 
    }
    \label{tab:mcs}
\end{table}

\section{Analysis}
\label{sec:analysis}

Data analysis was carried out similarly as described in Sec. 3.1 in \cite{ref:clouds}.
Namely, we analyzed the MC simulations of protons with the \texttt{ctapipe}\footnote{\url{https://github.com/cta-observatory/ctapipe}} version 0.12 and \texttt{lstchain}\footnote{\url{https://github.com/cta-observatory/cta-lstchain}} version 0.9.13 frameworks \citep{ctapipe, lstchain}. We reduced the raw data to the data level 1 (DL\,1, containing images of individual events in each telescope), using the \texttt{r0\_to\_dl1} script modified with stereo parameter calculation to generate the full package of DL\,1 parameters. This script allows us to 
perform image calibration and cleaning, as well as estimate Hillas, timing, leakage, concentration, and stereoscopic parameters from the simulated data. 

In the analysis, we apply a set of quality cuts to select well-reconstructed images carrying information about the additional absorption in a cloud. 
The cuts have been selected to limit the bias caused by the cloud.
For example, a commonly applied cut in the total light of the image would introduce a bias, as cloud-affected images are dimmer. 
The first stage of cuts are applied at the stereoscopic reconstruction level. 
We select only images with at least 20 pixels surviving the cleaning and composed of one island, which are reconstructed better.
We exclude images with the absolute value of time gradient\footnote{The time gradient describes the rate of change of the arrival time along the major axis of the ellipse.} less than 1 ns/m to avoid single muon-dominated images. 
We also exclude events with the centre of gravity lying outside of the cleaned image. 
All the events with at least two images surviving these criteria are kept for the stereoscopic reconstruction.

In the next step, for each image surviving the above-mentioned reconstruction we consider a second set of conditions.
To avoid nearby events, which have roundish images, without a clear height profile imprinted in them, we exclude events with the absolute value of time gradient below 5 ns/m. 
On the other hand, to exclude events with very large impact parameter, that are cut at the edge of the camera and do not follow the geometrical model, we only keep events with the absolute value of the time gradient below 15 ns/m. Cutting in time gradient is motivated by the fact that this parameter is very little affected by the cloud presence (see \citealp{ref:clouds}).
We compute that the expected rate of images surviving such cuts for a simulated array of LST prototype-like telescopes, obtained from the proton simulations, is $\sim60$~Hz. 
This allows us to monitor the atmospheric conditions even at short time scales of minutes. 
The absorption of light in clouds is biasing the selection of events, however, thanks to the careful selection of cut variables the rate of the selected images does not vary strongly with the atmospheric conditions. 
Namely, for the studied clouds the rate is only lowered by 10\% (for $T\approx 0.8$ cloud at 6.5 -- 8.5~km a.g.l) to 33\% (for $T\approx 0.4$ cloud at 6.5 -- 8.5~km a.g.l).

\begin{figure}[t!]
    \centering
    \includegraphics[width=0.45\textwidth]{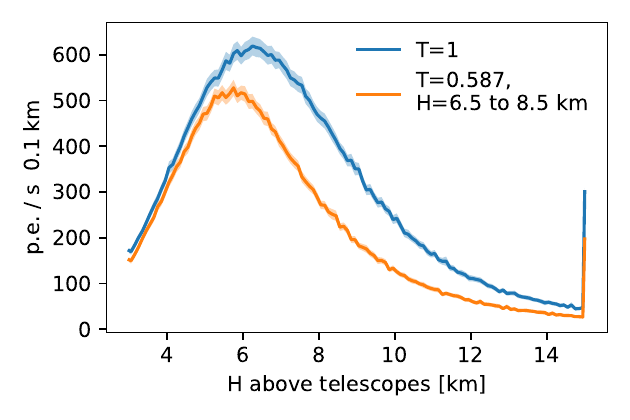}
    \caption{Aggregated longitudinal distribution (as a function of reconstructed emission height) of the emitted Cherenkov radiation reconstructed from proton simulations. 
    The blue line shows the case of a clear atmosphere $T=1$, while the orange line shows $T\approx 0.6$ cloud at 6.5 -- 8.5~km a.g.l. 
    The last bin contains the integrated emission above 15~km a.g.l. 
    The shaded region shows the expected uncertainty of the measured aggregated profile (binned every 100~m) assuming 5-minute long observations.
    }
    \label{fig:longi}
\end{figure}

An example of an aggregated longitudinal distribution for the baseline cloud, reconstructed according to the method presented in Section~\ref{sec:atmprof} is shown in Fig.~\ref{fig:longi}.
The small differences in the two distributions below the base of the cloud are due to the resolution of the method. 

We derived the expected statistical uncertainty of the aggregated profile by following the usual error propagation. 
Namely, we compute the square root of the sum over events of squared signal values associated with each height bin. 
The uncertainty will then scale with the inverse of the square root of observation time. 
It is sufficient to integrate 5~min long observations to achieve statistical uncertainty below 5\% up to the height of 12~km above the telescopes (see the shaded region in  Fig.~\ref{fig:longi}). 

\section{Results}\label{sec:results}

\begin{figure*}[t!]
    \centering
    \includegraphics[width=0.33\textwidth, trim=305 0 10 0, clip]{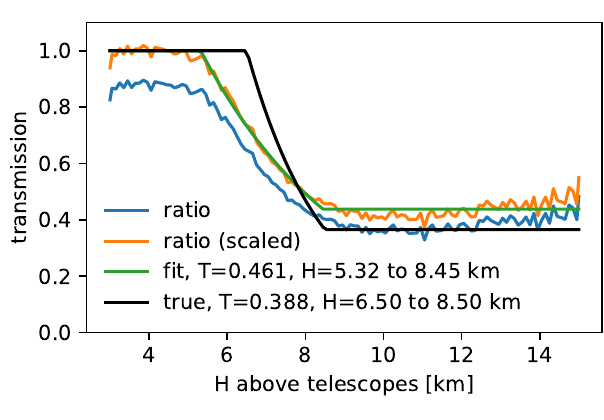}
    \includegraphics[width=0.33\textwidth, trim=305 0 10 0, clip]{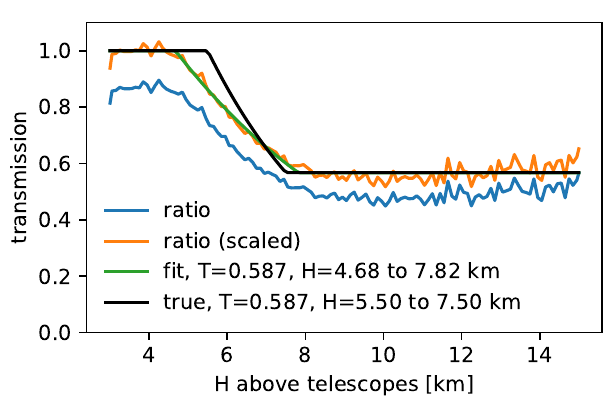}
    \includegraphics[width=0.33\textwidth, trim=305 0 10 0, clip]{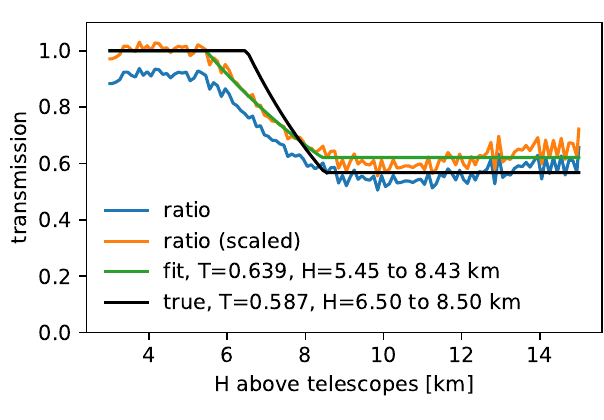}
    \includegraphics[width=0.33\textwidth, trim=305 0 10 0, clip]{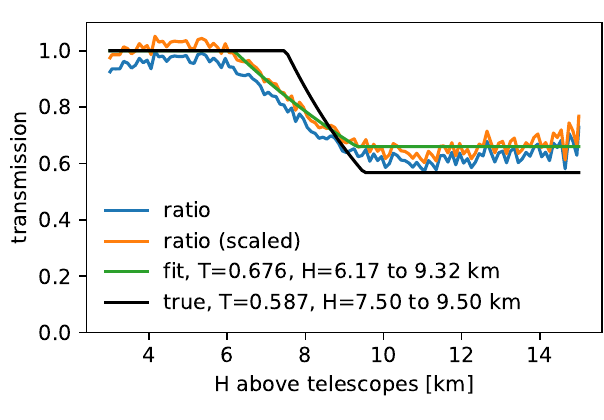}
    \includegraphics[width=0.33\textwidth, trim=305 0 10 0, clip]{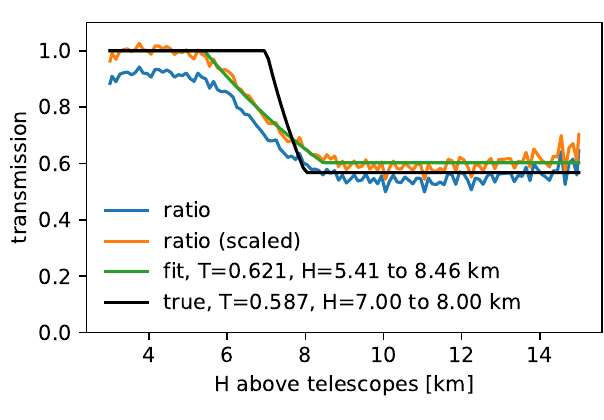}
    \includegraphics[width=0.33\textwidth, trim=305 0 10 0, clip]{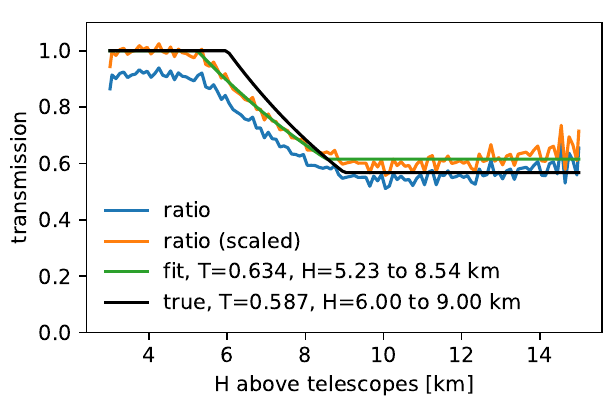}
    \includegraphics[width=0.33\textwidth, trim=305 0 10 0, clip]{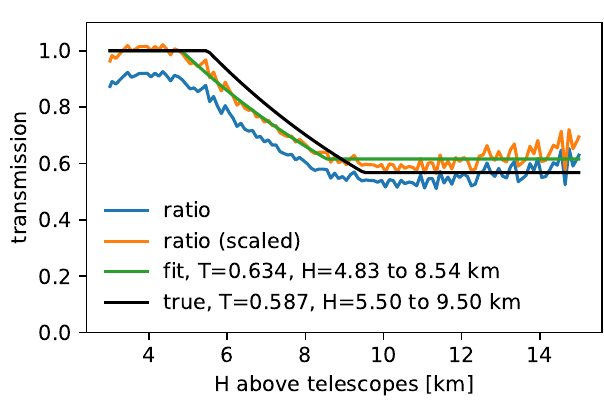}
    \includegraphics[width=0.33\textwidth, trim=305 0 10 0, clip]{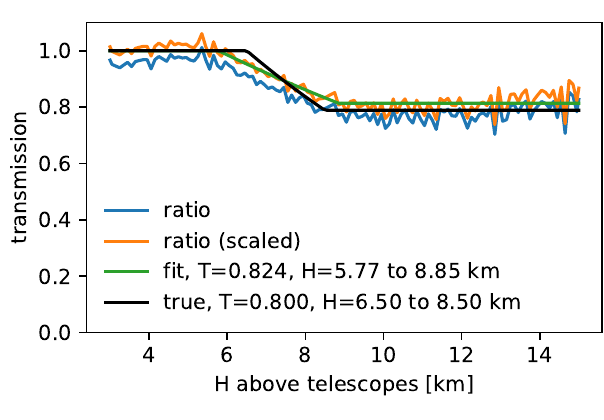}
    \caption{Ratios of the longitudinal distributions for different simulated clouds (see the description of transmission $T$ and height $H$ range in the last line of the legends in individual panels).
    The blue line shows the raw ratio, while the orange shows the ratio renormalised at 3 -- 4~km a.g.l.
    The black line shows the simulated atmospheric profile, while red is the reconstructed profile from the fit to the orange line. }
    \label{fig:ratios}
\end{figure*}

\begin{figure}[tp]
    \centering
    \includegraphics[width=0.49\textwidth]{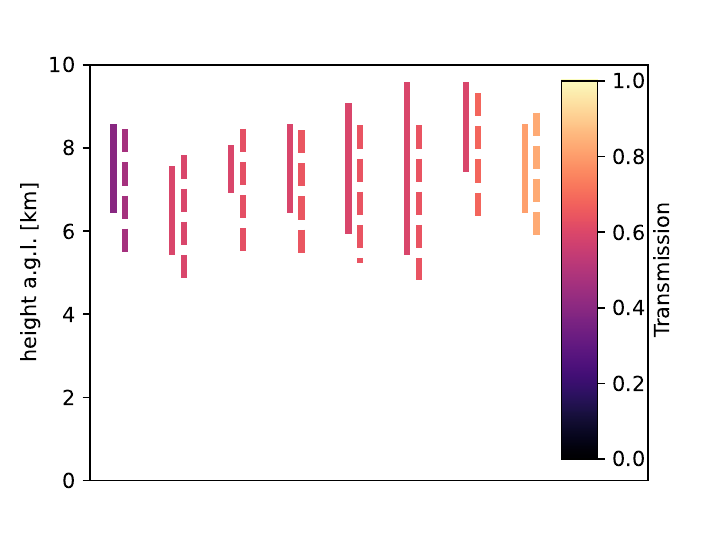}
    \caption{Bias in the reconstructed parameters of the cloud. Each pair of lines is one considered cloud with their true parameters shown in solid line and reconstructed in dashed. 
    The Y axis symbolises vertical extend of the cloud, while the color scale the total transmission. 
    }
    \label{fig:cloudpar_bias}
\end{figure}

For each simulated cloud, we compute the aggregated longitudinal distribution of the emitted light and calculate the ratio to the one obtained with a clear atmosphere. 
To counteract the residual selection bias in the number of registered photoelectrons introduced by the cloud, we normalise the derived ratio to 1 in the emission height range of 3 -- 4~km a.g.l., 
well below the height of the simulated clouds. 
The results are shown in Fig.~\ref{fig:ratios} and compared with the simulated transmission profiles of the cloud.
The difference of the estimated parameters of the cloud and the simulated values (i.e. the bias of the method) is summarised in Fig.~\ref{fig:cloudpar_bias}.
%

The method can reconstruct relatively well (within a few per cent of absolute accuracy) the total transmission of the cloud, typically slightly overestimating the transmission of the cloud.
In all the simulated cases there is a bias underestimating the height of the cloud.
The bias is mostly reconstructing the cloud base at lower heights, while the top of the cloud is reconstructed in most of the cases (except $T=0.587$, $H=5.5 - 9.5$~km) with an absolute bias lower than half a km. 
This results in a net bias of the geometrical centre of the cloud being reconstructed at lower heights, 
but such a bias is relatively small compared to the size of the simulated clouds. 
Due to the discussed earlier broadness of the angular offset distribution at a given emission height, there is a limitation in reconstructing the profile of the cloud. This causes a considerable overestimation of the thickness of the cloud. 
As a result, cloud structures narrower than $\sim 3$~km have overestimated geometrical thickness. 

\subsection{Effect of systematic uncertainties on the method}
\label{sec:syst}
The IACT observations are burdened with a number of systematic uncertainties. 
In this section, we discuss their possible effect on the atmospheric transmission profiles derived with the proposed method. 
As the proposed method requires a comparison of two data sets (cloud and a reference data set with a clear atmosphere), we pay particular attention to systematic uncertainties that might change in time or make such comparisons more difficult.  

\paragraph{Helium and higher elements}
While protons are the dominant background for the IACTs, the Cosmic Rays include also helium and higher elements.
Even while they are less prone to trigger IACTs, \citealp{2012APh....35..435A} estimated a $\sim20\%$ of additional rate of such events before the gamma/hadron separation. 
To evaluate if those higher elements would affect the proposed method, we simulated helium nuclei (the second most abundant element in Cosmic Ray spectra) for the case of cloudless ($T=1$) and the baseline cloud ($T= 0.587$ at $6.5-8.5$ km a.g.l.) following a power-law with spectral index of $-2$ in the energy range from 40~GeV to 1200~TeV.
The helium flux is normalised to 50\% of the proton flux and mixed with the proton simulations, then the proposed model is applied to the whole sample.
The helium images are about 14\% of all the images used in the analysis.

\begin{figure}[t!]
    \centering
    \includegraphics[width=0.4\textwidth]{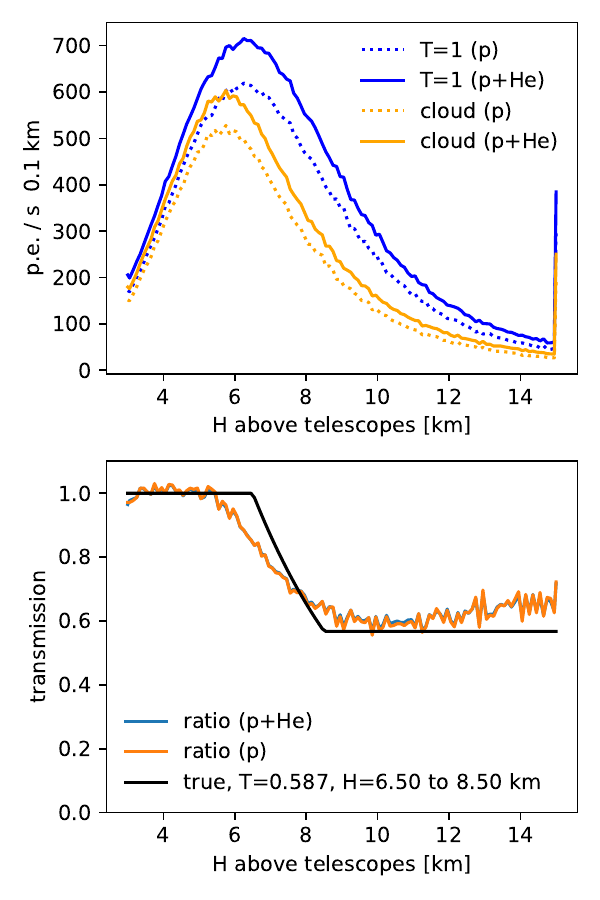}
    \caption{Top panel: aggregated longitudinal distribution of the emitted Cherenkov light from protons (dotted) or protons and helium (solid lines) for cloudless case (blue) and $T\approx 0.6$ cloud at $6.5-8.5$~km a.g.l. (orange)
    Bottom panel: ratios of these distributions for the sum of proton and helium (blue) or protons alone (orange) compared to the transmission profile of the simulated cloud (black). 
    Note that due to very small effect introduced by the Helium the blue line in the bottom panel is nearly covered by the orange one. } 
    \label{fig:helium}
\end{figure}

In Fig.~\ref{fig:helium} we compare the results of the proposed method applied to protons only or such a mixture of proton and helium images. 
As both observations without clouds and those in the presence of clouds capture showers from the identical cosmic ray composition, and considering that a majority of the images employed in the method originate from proton showers, the inclusion of helium nuclei in the simulations has a minimal effect on the reconstructed transmission profile of the cloud. 
Therefore, we conclude that also the effect of higher elements is negligible. 

\begin{figure}[t!]
    \centering
\includegraphics[width=0.49\textwidth]{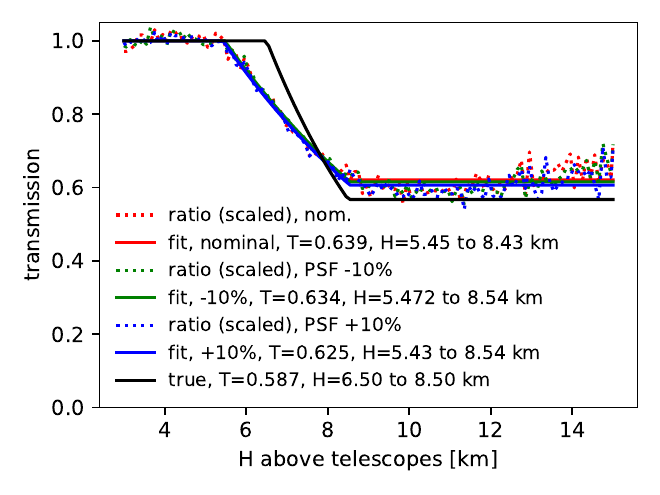}
\caption{Comparison of the transmission profiles derived using MC simulations with varying optical point spread function.
The true cloud profile is shown in black. The reconstructed profile for the nominal PSF value is shown in red, while the case of decreased/increased PSF values by 10\% is shown in the green/blue line.
The reconstructed transmission profiles are shown in dotted lines and their fits with solid lines. } \label{fig:psf}
\end{figure}

\paragraph{Optical PSF}
The amount of light registered by the telescope from a particular shower depends not only on the atmospheric transparency but also on the conditions of the telescope itself. 
The optical Point Spread Function (PSF) of Cherenkov telescopes can degrade slightly with time (the degradation is normally counteracted with periodic maintenance activities). 
The expected changes of the optical PSF are however small (in particular when compared to the angular extend of the large showers used in this method), therefore the performance of the presented method is not expected to be degraded by them. 
However night-to-night variations in the optical PSF result in its value in the reference cloudless data set being slightly different than the one in the data from which we want to derive the cloud transmission profile. 
According to CTA requirement B-TEL-0135, the optical PSF of the telescopes must be known with an accuracy better than 10\%. 
Therefore, we perform dedicated simulations of a cloud in which the individual mirror optical PSF of the telescopes is modified by $\pm10\%$.
Next, we compare such simulations with the cloudless simulations with nominal values of the optical PSF (see Fig.~\ref{fig:psf}).
The proposed method is not sensitive to such optical PSF changes. 
The rate of the images selected for the analysis varies by less than 1\%. 
Also the relative difference in the reconstructed transmission is only 1 -- 2\%. 
This simplifies the practical applications of the method in terms of finding reference cloudless data set appropriate for evaluating cloud-affected data. 

\paragraph{Mirror reflectivity}
Similarly to the optical PSF, the reflectivity of the mirrors also evolves in time (but also the transmission of e.g. camera protection window) due to e.g. dust deposits or degradation with time. 
According to the CTA requirement A-PERF-2050, the absolute throughput of the telescopes should have a systematic uncertainty lower than 8\%. 
To test the effect of such a systematic effect on the proposed method, we generate MC simulations with an 8\% difference in reflectivity value compared to the nominal one. 
The calculated rate of the selected images for such Monte Carlo simulations is also about 8\% different than for nominal ones.  

\begin{figure}[t!]
    \centering
\includegraphics[width=0.49\textwidth]{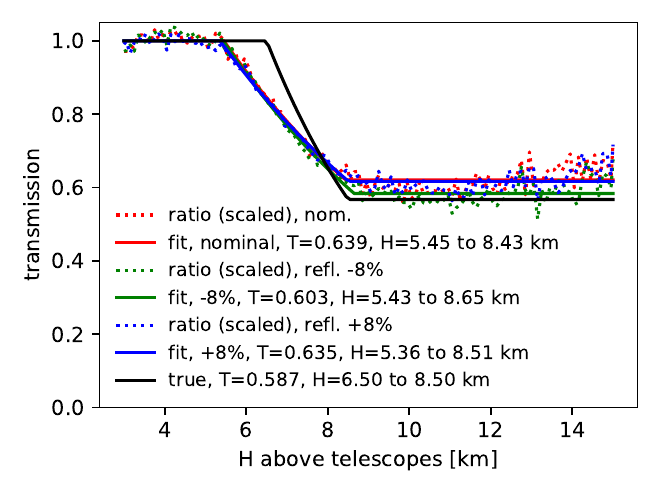}
\caption{Comparison of the transmission profiles derived using MC simulations with varying total reflectivity.
The true cloud profile is shown in black. The reconstructed profile for the nominal value of the reflectivity is shown in red, while the reflectivity decreased and increased by 8\% is shown in green and blue respectively.
The reconstructed transmission profiles are shown in dotted lines and their fits with solid lines. }
    \label{fig:mdr}
\end{figure}

We then construct the ratio of the aggregated longitudinal distribution of the cloud with modified reflectivity to the clear sky simulations with nominal reflectivity. 
The resulting transmission profiles and their fits are shown in Fig.~\ref{fig:mdr}.
Even such a rather large difference in the telescope reflectivity does not affect the proposed method considerably. 
The relative variations in the reconstructed transmission ($\sim 6$\%) are smaller than the change applied in the mirror reflectivity.

\paragraph{Pointing direction (zenith distance)}
The rate of events and distribution of the detected Cherenkov light depends on the pointing direction of the telescopes. 
In particular, the properties of the showers depend on the zenith distance angle, $\theta$ of the observations and therefore the cloud and reference samples should be taken at similar values. 
To investigate it we generated proton simulations with cloudless and reference cloud conditions 
for three additional zenith distance angles: $5^\circ$, $45^\circ$ and $60^\circ$. The maximum impact parameter for this set of simulations is calculated following \cite{Abe2023PerformanceData} approach where nominal $r_\mathrm{max}$ for $\theta=20^\circ$ is folded by a factor of $\cos^{-0.5}\theta$.

\begin{figure}[t!]
    \centering
\includegraphics[width=0.49\textwidth]{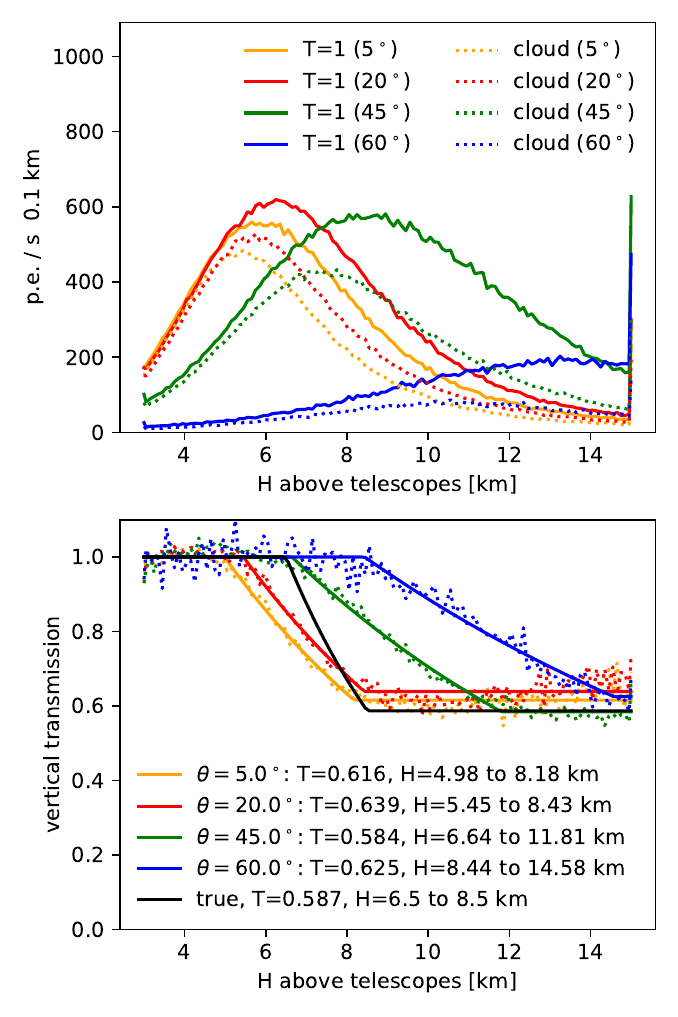}
\caption{Top panel: comparison of the aggregated longitudinal distributions for cloudless case (solid) and $T\approx 0.6$ cloud at 6.5 -- 8.5 km a.g.l. (dotted) for different zenith distance angles of the observations: $5^\circ$ (orange), $20^\circ$ (red), $45^\circ$ (green) and $60^\circ$ (blue).
Bottom panels: the corresponding reconstructed vertical transmission (dotted lines) and the fit to it (solid lines) compared to the simulated one (black line). 
The colours in the bottom panel correspond to the same zenith distance angles as in the top panel.
}
    \label{fig:zenith}
\end{figure}

In Fig.~\ref{fig:zenith} we present the obtained distribution of the registered emission heights and reconstruction of the cloud for those higher zenith distance angle observations, compared to the $\theta=20^\circ$ case considered before.
The reconstructed distribution of the emission height shows a strong dependence on the zenith distance angle. 
This is caused by the thickness of the atmosphere increasing with the zenith distance angle, resulting in the shower developing higher. The obtained images are also more compact images and the performance of the stereoscopic reconstruction suffers from the shrunken baseline between the telescopes. 
We find that as long as the cloud-affected observations are compared with the reference cloudless observations at the same zenith angle, the proposed method allows the reconstruction of the transmission of the cloud with similar accuracy for all the considered pointing angles.  
However, for higher zenith distance angles, the geometrical thickness of the cloud, as well as its centre is overestimated. 

In addition to the degradation of the method accuracy for observations at higher zenith distance angle, 
the method requires  reference cloudless observations at all relevant zenith distance angles.  
To counteract this effect, for the possible cases when finding a cloudless reference sample at the same zenith is not feasible, we investigate a possible scaling procedure of the $T=1$ aggregated height distribution of the registered light.
Let us assume that the reference observations are taken at a zenith distance angle of $\theta_0$ with the corresponding height distribution of the emitted light $M(h_0;\theta_0)$.
The objective is to scale the distribution to the zenith distance angle  $\theta_c$, at which the cloud-affected data are taken, i.e. to obtain $M'(h_c;\theta_c)$.
We convert the heights assuming that the emission is dependent only on the total thickness of the atmosphere $D(h)$ (in units of $\mathrm{g\,cm^{-2}}$), therefore:
\begin{equation}
    D(h_c) / \cos\theta_c = D(h_0) / \cos\theta_0. \label{eq2}
\end{equation}

The observations at different zenith distance angles would result also in the modification of the density of Cherenkov photons reaching the telescope. 
Namely, the density should scale with the inverse of the area over which the photons are distributed, which can be estimated as $A=\pi (\alpha(h) h / \cos\theta))^2$, where $\alpha$ is the height-dependent Cherenkov angle. 
However, the effective collection area of the telescope would scale linearly with $A$, cancelling out that effect in the aggregated emitted light distribution. 
Therefore, the distribution of $M'(h_s)$ must be only corrected for the difference of binning of the emission height induced by Eq.~\ref{eq2}:
\begin{equation}
    M'(h_c;\theta_c) = M(h_0;\theta_0) \Delta h_0 / \Delta h_c. \label{eq3}
\end{equation}

\begin{figure*}[t]
    \centering
    \includegraphics[width=0.49\textwidth]{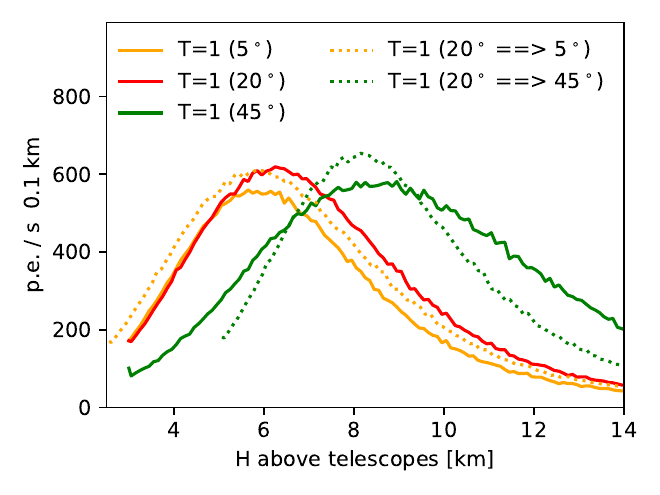}
    \includegraphics[width=0.49\textwidth]{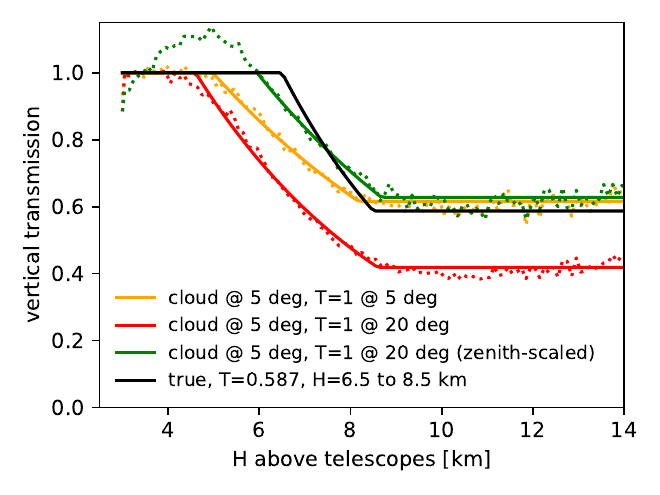}
    \caption{Left panel: scaling of aggregated longitudinal distribution. The simulated cloudless distributions for zenith distance angles of $5^\circ$, $20^\circ$ and $45^\circ$ are shown with solid orange, red and green curves respectively.
    The dashed lines show the distributions scaled following Eq.~\ref{eq2} and Eq.~\ref{eq3}.
    Right panel: reconstructed vertical transmission.
    The orange curve shows the case of matching zenith angle distance between cloud-affected and cloudless data.
    Red and green curves show the case of a mismatched zenith angle distance without and with the scaling of Eq.~\ref{eq2} and Eq.~\ref{eq3}. 
    The true transmission curve is shown in black. 
    }
    \label{fig:zd5_corr}
\end{figure*}

In the left panel of Fig.~\ref{fig:zd5_corr} we compare such obtained scaled distribution with the one derived from the full simulations. 
Such simple scaling is able to reproduce closely the peak position of the distribution. 
For a small difference in cosine of the zenith 
the complete distribution is reproduced well
, except for a small mismatch in normalisation that is irrelevant to the proposed method. 
It should be noted that the simple scaling does not take into account a number of effects, in particular the dependence of the Cherenkov light absorption on the zenith and modification of the images by zenith angle that would influence the selection cuts. 

In the right panel of Fig.~\ref{fig:zd5_corr} we test the accuracy of such zenith-scaling in the proposed method.
While using matching zenith angle the reconstructed transmission is $0.616$, using $\theta_0=20^\circ$ cloudless simulations for  $\theta_c=5^\circ$ cloud-affected simulations causes a large bias in the estimated transmission ($0.419$).
The scaling of the reference distribution using Eq.~\ref{eq2} and Eq.~\ref{eq3} removed this bias (the obtained transmission is $0.628$), however small $\pm10\%$ distortion of the reconstructed transmission curve at low heights is present.

\begin{figure}[t!]
    \centering
\includegraphics[width=0.49\textwidth]{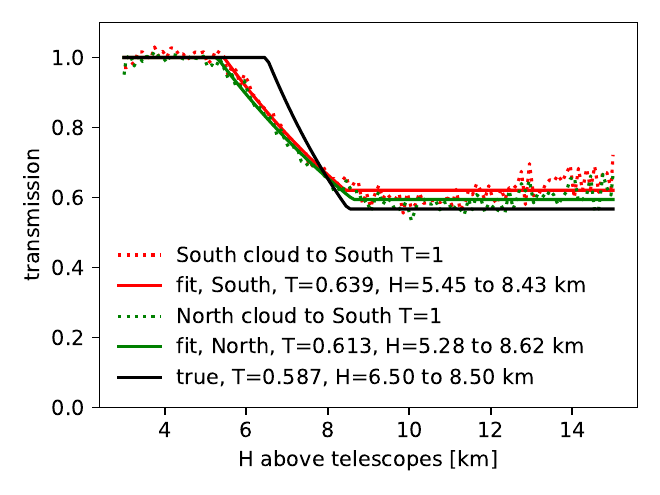}
\caption{
Influence of pointing direction on the reconstructed cloud transmission profile. 
The true cloud profile is shown in black. 
The reconstructed profile for the case of matching pointing between cloudless and cloud observations is shown in red.
In the case in which different pointing is used for a reference, cloudless observations are shown in green. 
The reconstructed transmission profiles are shown in dotted lines and their fits with solid lines. 
}
    \label{fig:azimuth}
\end{figure}

\paragraph{Azimuth angle}
Even in the case of low-zenith observations, the geomagnetic field can affect the showers introducing an azimuth dependence (see e.g. \citealp{1992JPhG...18L..55B,2013APh....45....1S}).
To test this effect, we compare the performance of the method when the ratio is constructed from matching pointings of the telescope (both cloudless and cloud cases have telescopes pointing to the South) with the mismatching case (cloudless simulations have telescopes pointing South, while cloud is observed when pointing North). 
The results for $20^\circ$ zenith distance angle simulations are shown in Fig.~\ref{fig:azimuth}.
The mismatch of the azimuth direction (at the same zenith distance angle) has a small effect on the method.
The absolute difference in the reconstructed cloud transmission is $\sim3\%$. 

\begin{figure}[t!]
    \centering
\includegraphics[width=0.49\textwidth]{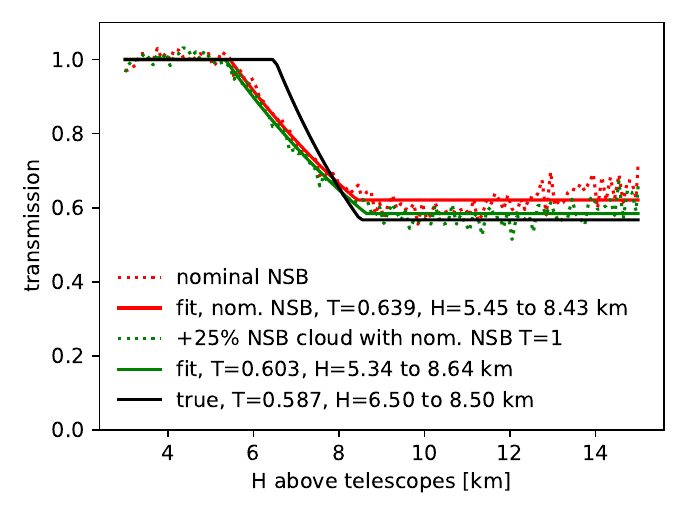}
\caption{
Influence of NSB level on the reconstructed cloud transmission profile. 
The true cloud profile is shown in black. 
The reconstructed profile for the case of nominal NSB for both cloudless and cloud observations is shown in red.
The case in which cloud-affected observations are simulated with 25\% higher level of NSB than the reference cloudless observations is shown in green. 
The reconstructed transmission profiles are shown in dotted lines and their fits with solid lines. 
}
    \label{fig:nsb}
\end{figure}

\paragraph{Night Sky Background (NSB)}
When comparing the cloud-affected and reference (cloudless) observations, the two observations might not cover the same field of view. 
This introduces another possible systematic error in the method if e.g. dimmer extragalactic field is compared with a more luminous Galactic path of a sky. 
To quantify the possible effect of such a mismatch, we generated a dedicated production of the baseline cloud with NSB increased by 25\%\footnote {While in reality, the increase of the NSB level would also require an increase of trigger thresholds, we neglect this effect in our simulations, since the proposed method is applied only to larger events that are easy to trigger.}.
The results of applying the method with an NSB mismatch between two simulation sets are presented in Fig.~\ref{fig:nsb}.
The absolute difference of the reconstructed transmission depending on the NSB level in the reference simulations is at the level of 4\%. 
Therefore slight mismatches of the NSB level between the two datasets are not expected to introduce a large systematic error in the proposed method. 

\section{Conclusions and discussions}\label{sec:summary}

We have developed a method to evaluate the transmission profile (height-dependent transmission) of a cloud using data recorded by IACTs. 
The method is based on observations of abundant proton-initiated showers, that allow monitoring of the atmospheric conditions on time scales of the order of minutes. 
We exploited simulations of four LST telescopes, that are planned to become part of the CTAO,  to evaluate the performance of the method.
The aggregated profiles of the emitted Cherenkov light can be reconstructed with a statistical accuracy better than 5\% with just a 5-minute long exposure.
The total vertical transmission of the cloud is reconstructed remarkably well for a range of simulated clouds, typical for the CTA-North location.
While the height of the cloud is slightly biased, for low zenith observations the bias does not exceed $\sim$0.8~km. 
Typically the heigh of the top of the cloud is better reconstructed than the height of its base. 
The geometrical thickness of the cloud is however more poorly estimated, particularly for clouds thinner than 3~km.
Due to residual biases and the need for normalisation, the method is not directly applicable to absorption occurring at very low heights, such as in the case of Saharian Air Layer (the so-called calima, see e.g. \citealp{2011ACP....11.6663R}).  

We tested a number of possible systematic errors that might affect the method when applied to real data.
We positively validated the assumption that it is sufficient for the method to consider only proton-initiated showers, by showing that helium-initiated showers have a negligible effect on it.
Moreover, we also showed that small changes in the telescope's optical performance (reflectivity or optical PSF) have very little effect on the method.
Also, the azimuth dependence (mainly caused by the geomagnetic field effect) and the background light level have little effect on the reconstructed cloud parameters.

Due to the strong dependence on the distribution of the emission heights of the registered Cherenkov light, the method must compare the cloud and cloudless observations taken at the same zenith angles. 
For the cases when this is not feasible, we proposed a scaling method that significantly lowers the bias induced by the mismatch of the zenith distance angles.
Additionally, for high zenith observations, while the method reconstructs the total transmission correctly, it provides poorer accuracy in both the geometrical centre and the thickness of the cloud. 

The described method, with a typical statistical and systematic accuracy of a few per cent, has great potential in practical application.
While it cannot substitute a LIDAR measurement in all cases, it provides independent, continuous monitoring of the cloud conditions without the need for any additional device or dedicated observations.
Instead, the information about the cloud conditions can be extracted from the copious isotropic background events present in the science-driven observations of gamma-ray sources. 
While the performance of the method was evaluated on the example of an array of four LST telescopes, it can be used with any stereoscopic array of Cherenkov telescopes with imaging capability.
The information derived by this method is then universal, as it can be used in the correction of the data using a number of different methods (e.g. \citealp{Schmuckermaier23e,ref:clouds}) and/or as a measure of the quality of the atmosphere.

\section*{Acknowledgements}
This work is supported by Narodowe Centrum Nauki grant number 2019/34/E/ST9/00224 and the Croatian Science Foundation grant IP-2022-10-4595.
This work was conducted in the context of the CTA LST Project. 
The authors are grateful to CTA and LST Consortia internal reviewers: F. Di Pierro, F. Schmuckermaier, J. L. Contreras and O. Blanch for their comments which helped to shape the paper.





\end{document}